\documentclass[twocolumn,prb,showpacs,floats]{revtex4}
\usepackage[dvips]{graphicx}
\usepackage{dcolumn}
\usepackage{color}

\usepackage{multirow} 

\usepackage{bm}

\newcommand{\SCO}{Sr$_2$CoO$_4$}
\newcommand{\SLCO}{Sr$_{1.5}$La$_{0.5}$CoO$_4$}

\begin{document}

\title{Metal-insulator transition in Sr$_{2-x}$La$_x$CoO$_4$ driven by spin-state transition}
\author{Hua Wu}
\thanks{Email address: wuh@fudan.edu.cn}
\affiliation{Laboratory for Computational Physical Sciences (MOE), State Key Laboratory of Surface Physics, and Department of Physics, Fudan University, Shanghai 200433, China}

\date{today}

\begin{abstract}

We sought the origin of the metal-insulator transition in Sr$_{2-x}$La$_x$CoO$_4$, 
using electron-correlation corrected density functional calculations. 
Our results show that Sr$_2$CoO$_4$ is in an intermediate-spin (IS, $t_{2g}^4e_g^1$) state
and a strong Co$^{4+}$ $3d$-O $2p$ hybridization is responsible for its ferromagnetic metallicity. 
Upon La doping, however, a spin-state transition occurs in 
Sr$_{1.5}$La$_{0.5}$CoO$_4$: IS Co$^{4+}$$\times$2 + 1$e$ $\rightarrow$ LS 
Co$^{4+}$ ($t_{2g}^5$) + HS Co$^{3+}$ ($t_{2g}^4e_g^2$) (LS: low spin; HS: high spin). 
Then the spin-state transition suppresses an electron hopping via a spin-blockade
and gives rise to the insulating behavior of Sr$_{1.5}$La$_{0.5}$CoO$_4$. 
A corresponding superexchange accounts for its ferromagnetism.
Thus, spin state could provide a way to tune materials properties. 

\end{abstract}

\pacs{71.30.+h, 75.30.-m, 71.20.-b, 71.70.-d}

\maketitle
\section{Introduction}

It is quite common that a ferromagnetic (FM) material is metallic but an antiferromagnetic 
one is insulating (or semiconducting). Therefore, either a FM insulator or an 
antiferromagnetic metal seems 
to be an exception and would be of interest. In electron-correlated transition-metal 
oxides, charge, spin, orbital, and lattice degrees of freedom often couple to one another 
and result in abundant electronic and magnetic properties.\cite{Tokura} As a result, 
exceptions from the above `rule' emerge more than rarely.\cite{Braden} In this work, 
we will study such an exception --- the FM semiconducting cobaltate {\SLCO}, and the end 
material of the Sr$_{2-x}$La$_x$CoO$_4$ series, {\SCO} for a comparison.

Sr$_{2-x}$La$_x$CoO$_4$ is a group of interesting materials, which has the K$_2$NiF$_4$-type 
layered structure. {\SCO} is a FM metal,\cite{Matsuno,WangXL} and it could 
even be a half metal.\cite{Pandey} Upon La doping ($x$=0.5), {\SLCO} turns semiconducting 
but remains FM.\cite{Shimada,Chichev} With La doping up to $x$ = 1, SrLaCoO$_4$ becomes 
a paramagnetic insulator.\cite{Shimada,Chichev,Wu10} It is well known that cobaltates 
often have a spin-state issue.
The spin state depends on Hund exchange, crystal field, and band hybridization, and it has 
a strong impact on the magnetic and electronic properties of cobaltates. It is this 
spin-state issue that has brought about a lot of debates in literature. Partially because 
of this, Sr$_{2-x}$La$_x$CoO$_4$ draw much attention very recently.\cite{Hollmann08,Cwik,
Chang,Wu09,Helme,Tealdi,Babkevich,Hollmann11,Merz}

\begin{figure}[t]
\centering \includegraphics[width=6cm]{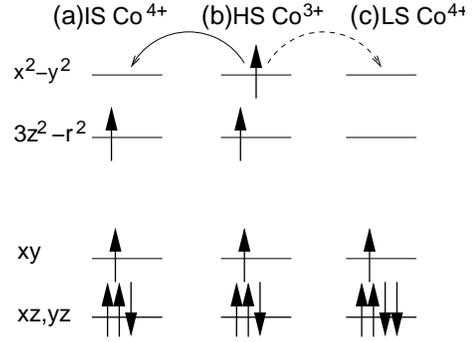} 
\caption{
Electronic configuration of (a) the IS and (c) LS state of a Co$^{4+}$ ion in a $c$-axis 
elongated CoO$_6$ octahedron (a tetragonal crystal field), and of (b) the HS state of a 
Co$^{3+}$ ion. While the $x^2$--$y^2$ electron would readily hop from the HS Co$^{3+}$ to
IS Co$^{4+}$ via Zener's double exchange, a hopping from the HS Co$^{3+}$ to LS Co$^{4+}$
is suppressed by a spin blockade. See more in the main text.
}
\end{figure}

In this work, using generalized gradient approximation plus 
Hubbard $U$ (GGA+$U$)\cite{Anisimov} calculations, we seek the origin of 
the metal-insulator transition induced by La doping 
in {\SCO}, and pay attention to the FM semiconducting behavior of {\SLCO}.
In Fig. 1, we sketch the Co $3d$ crystal-field levels in Sr$_{2-x}$La$_x$CoO$_4$ 
with a $c$-axis elongated CoO$_6$ 
octahedron. The electronic configurations are shown for an 
intermediate-spin (IS) state [Fig. 1(a)] and for a low-spin (LS) state [Fig. 1(c)] both 
relevant to a high-valent Co$^{4+}$ ion. Our following GGA+$U$ calculations find that 
Co$^{4+}$ in {\SCO} favors the IS state, though somewhat modified compared to 
Fig. 1(a) due to hybridization effects and on-site Coulomb interactions, and that 
a strong Co $3d$-O $2p$ hybridization results in FM metallicity of the compound. 
In Fig. 1(b), we show a high-spin (HS) state for a Co$^{3+}$, which emerges upon  
La doping. Here the Co$^{3+}$ HS state is conceived, based not only on 
our very recent work about SrLaCoO$_4$,\cite{Wu10} but also on a crystal-field scenario 
(see below). As SrLaCoO$_4$ has a Co$^{3+}$ HS-LS mixed 
state,\cite{Wu10,WangJ} less La-doping in {\SLCO} will most probably change the LS Co$^{3+}$
into a LS Co$^{4+}$ and leave the HS Co$^{3+}$ unchanged, in order to maximize Hund 
exchange. Then {\SLCO} would have the HS Co$^{3+}$ and LS Co$^{4+}$ ions. In this sense, 
upon La doping in {\SCO}, there would be an IS-LS transition of the Co$^{4+}$ ions, besides
an introduction of the HS Co$^{3+}$ ions. As seen below, our calculations indeed confirm this spin-state transition and find it to be the origin of the metal-insulator transition.       

\section{Calculations and Discussion}

\begin{figure*}[t]
\centering \includegraphics[width=16cm]{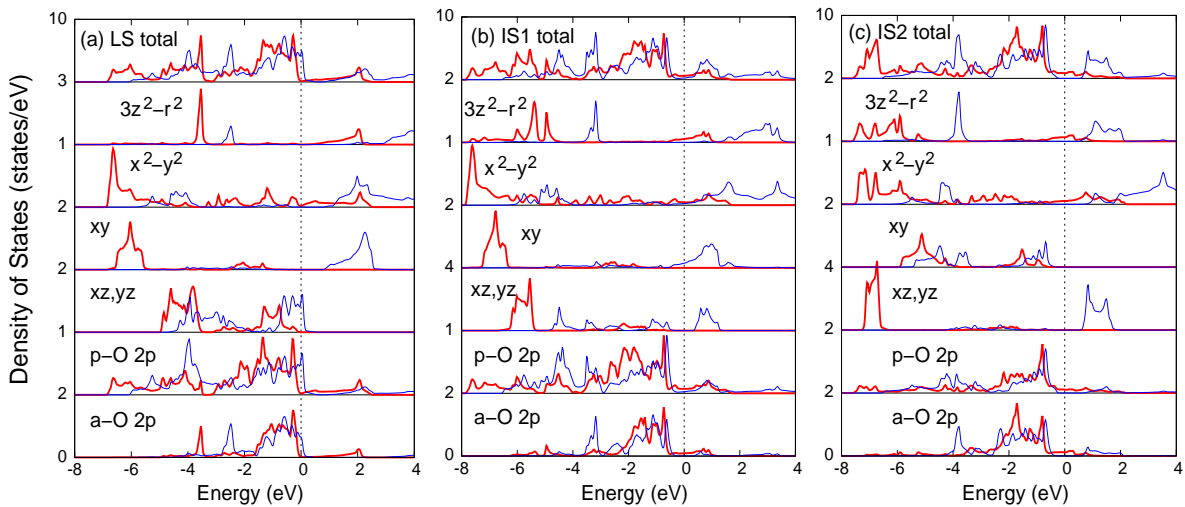} 
\caption{(Color online)
Total density of states (DOS), Co $3d$-orbital resolved DOS, planar O (p-O) 
and apical O (a-O) $2p$ DOS of FM {\SCO} in (a) LS, (b) IS1, and (c) IS2 states
calculated by GGA+$U$. The red (bold) curves stand for the up spin, and 
the blue (solid) for the down spin. Fermi level is set at zero energy.
While the LS and IS1 states are metallic, the IS2 state is half metallic.  
The IS2 is the ground state, see also Table I.
}
\end{figure*}

We carried out GGA+$U$ calculations, using 
the full-potential augmented plane wave plus local orbital
code (Wien2k).\cite{Blaha} The structural data were taken from 
Refs.~\onlinecite{WangXL} and \onlinecite{Chichev}.
The muffin-tin sphere radii were chosen to be 2.5, 2.0, and 1.5 Bohr for (Sr,La), Co, and O
atoms, respectively. A virtual atom with the atomic number $Z$=38.25 
(0.75$Z_{\rm Sr}$+0.25$Z_{\rm Y}$) was used for the (Sr$_{1.5}$La$_{0.5}$) sites, as
Sr and La(Y) ions are in most cases simple electron donors. 
The plane-wave cut-off energy of 16 Ry was set for the interstitial wave 
functions, and 1200 (600) {\bf k} points for integration over the Brillouin zone 
of {\SCO} (of {\SLCO} with a doubled unit cell). The spin-orbit coupling was included by 
the second-variational method with scalar relativistic wave functions. 
The typical value of $U$=5 eV and Hund exchange of 0.9 eV were used
in our GGA+$U$ calculations to account for the electron correlation of the 
Co $3d$ states.\cite{Wu10,WangJ} 

\subsection{Sr$_2$CoO$_4$}

We first calculated {\SCO}. As seen in Table I, our configuration-state constrained GGA+$U$
calculations find three stable FM solutions, one LS and two 
IS states. Indeed, the IS states are more stable than the LS state.  
The LS state is metallic, see Fig. 2(a). The up-spin $t_{2g}$ ($xy$ and $xz$/$yz$) orbitals
are fully occupied, and the down-spin $xz$/$yz$ are almost fully filled. Note that owing to 
the high valence of the Co$^{4+}$ ion and its negative charge-transfer energy,\cite{Potze} 
a strong $pd\sigma$ hybridization with the ligand oxygens bring about a large occupation 
on the $e_g$ ($3z^2$--$r^2$ and $x^2$--$y^2$) orbitals. This accounts for the calculated
total spin moment of 2.03 $\mu_B$/fu which is increased from the formal LS $S$=1/2 state. 
The Co ion has a local spin (orbital) moment of 1.82 (0.09) $\mu_B$ within its muffin-tin sphere,
see Table I.

\begin{table}[b]
\caption{ 
The relative total energies $\Delta$$E$ (meV/fu) of FM {\SCO} (SCO) in one LS and two IS states
calculated by GGA+$U$, the total spin moment $M$ and the local spin/orbital moment of Co ions 
in unit of $\mu_B$. A reference to figures is included. The corresponding data are listed 
for FM {\SLCO} (SLCO). The Co$^{3+}$ HS state is marked bold for clarity. Note that in our GGA+$U$
calculations, the IS Co$^{4+}$-IS Co$^{3+}$ state converges to the LS-HS ground state.
}
\begin{tabular}{l@{\hskip3mm}r@{\hskip3mm}c@{\hskip3mm}c@{\hskip3mm}c@{\hskip3mm}c} 
\hline\hline
SCO  & $\Delta$$E$ & $M$ ($\mu_B$/fu) & Co$^{4+}$ $m_s$/$m_o$& & Fig. \\ \hline
LS & 0 & 2.03 & 1.82/0.09 & & 2(a) \\
IS1 & --56 & 2.79 & 2.53/1.02 & & 2(b) \\
IS2 & --155 & 3.00 & 2.61/0.10 & & 2(c) \\ \hline\hline
SLCO  & $\Delta$$E$ & $M$ ($\mu_B$/2fu) & Co$^{4+}$ & Co$^{3+}$ & Fig. \\ 
\hline
IS-{\bf HS} & 0 & 7.00 & 2.85/0.13 & 2.89/0.85 & 3 \\
LS-{\bf HS} & --139 & 5.00 & 1.70/0.01 & 2.88/0.84 & 4 \\
\hline\hline
\end{tabular}
\end{table}

Starting from the electronic configuration 
$t_{2g\uparrow}^3$($3z^2$--$r^2$)$_{\uparrow}^1$($xz$+$iyz$)$_{\downarrow}^1$ for the IS1 state, 
we also get a metallic solution with a total spin moment of 2.79 $\mu_B$/fu. 
The up-spin $e_g$ bands
get almost doubly occupied due to the strong $pd\sigma$ covalency, and they cross Fermi level
together with the down-spin $xy$ band, see Fig. 2(b). 
The down-spin $xz$+$iyz$ ($Y_{21}$) orbital is fully occupied and
well separated from the unoccupied $xz$--$iyz$ ($Y_{2-1}$) orbital. 
It is the $xz$+$iyz$ electron that
contributes most to the calculated orbital moment of 1.02 $\mu_B$. The local spin 
moment is 2.53 $\mu_B$ for the IS1 Co ion. 

The most stable IS state is achieved self-consistently from the initial configuration state
$t_{2g\uparrow}^3$($3z^2$--$r^2$)$_{\uparrow}^1$$xy_{\downarrow}^1$. 
It is FM half metallic, see Fig. 2(c).
Only the up-spin $e_g$ bands cross Fermi level. This solution has an integer spin moment
of 3 $\mu_B$/fu as expected for the formal IS $S$=3/2 state. The Co$^{4+}$ ion has a local
spin (orbital) moment of 2.61 (0.10) $\mu_B$, see Table I. This IS2 state is more stable
than the LS state by 155 meV/Co and than the IS1 state by 99 meV/Co. As the $xy$
singlet is a higher level than the $xz$/$yz$ doublet in an elongated tetragonal
crystal field, it seems a bit surprising that the IS2 state is more stable than the IS1 state.
However, for the almost doubly occupied $e_g$ orbitals [Figs. 2(b) and 2(c)],      
the planar $x^2$--$y^2$ orbital is itinerant but the $3z^2$--$r^2$ orbital is relatively 
localized. As such, a stronger Coulomb repulsion between the $3z^2$--$r^2$ and $xz$/$yz$
electrons than that between $3z^2$--$r^2$ and $xy$ makes the filling of the down-spin 
$xy$ orbital energetically more favorable than that of the down-spin $xz$/$yz$.  
Apparently, the FM half-metallic solution of {\SCO} is due to the significant $pd\sigma$
hybridization between the IS Co$^{4+}$ ions and the planar oxygens. This is because 
the high-valent Co$^{4+}$ ion has a negative charge transfer energy\cite{Potze} 
and thus its actual 
configuration is more like Co$^{3+}$$\underline{L}$. Indeed, Fig. 2(c) shows almost the
$3d^6$ state of the HS Co$^{3+}$ like and the ligand O 2$p$ hole states.   
Moreover, the planar-O $2p$ hole state is more itinerant than the apical-O $2p$. 
As a result, the calculated spin moment of 0.03 $\mu_B$ on the planar oxygen is smaller
than that of 0.10 $\mu_B$ on the apical oxygen. 

\subsection{Sr$_{1.5}$La$_{0.5}$CoO$_4$}

Now we turn to the calculations for the Co$^{4+}$-Co$^{3+}$ mixed-valent {\SLCO}.
As Co$^{3+}$ ion has a larger radius than Co$^{4+}$, longer Co$^{3+}$-O distances
would produce a weaker crystal field for Co$^{3+}$, compared with Co$^{4+}$.
In this sense, Co$^{3+}$ could be in a higher spin state than Co$^{4+}$.
Therefore, we first tested the IS Co$^{4+}$-HS Co$^{3+}$ state as a candidate for
{\SLCO}. As seen in Figs. 1(a) and 1(b), the $x^2$--$y^2$ electron would readily
hop from the HS  Co$^{3+}$ to IS  Co$^{4+}$, giving rise to a FM half-metallic
behavior via Zener's double exchange. This picture is indeed supported by our calculations,
see Fig. 3: the up-spin $x^2$--$y^2$ bands both of the Co$^{4+}$ and Co$^{3+}$ ions
are almost fully occupied but cross Fermi level due to the $x^2$--$y^2$ electron
hopping and the strong planar $pd\sigma$ hybridization. This FM half-metallic solution
has a total integer spin moment of 7 $\mu_B$/2fu as expected for the 
IS Co$^{4+}$-HS Co$^{3+}$ FM state ($S$=3/2 plus $S$=2). The formal HS Co$^{3+}$ and IS 
Co$^{4+}$ ions have, mainly due to the $x^2$--$y^2$ electron hopping, almost the same 
local spin moment (being about 2.9 $\mu_B$ each), see Table I. 
Note, however, that this half-metallic solution disagrees with the experimental semiconducting
behavior.\cite{Shimada,Chichev}

\begin{figure}[t]
\centering \includegraphics[width=8cm]{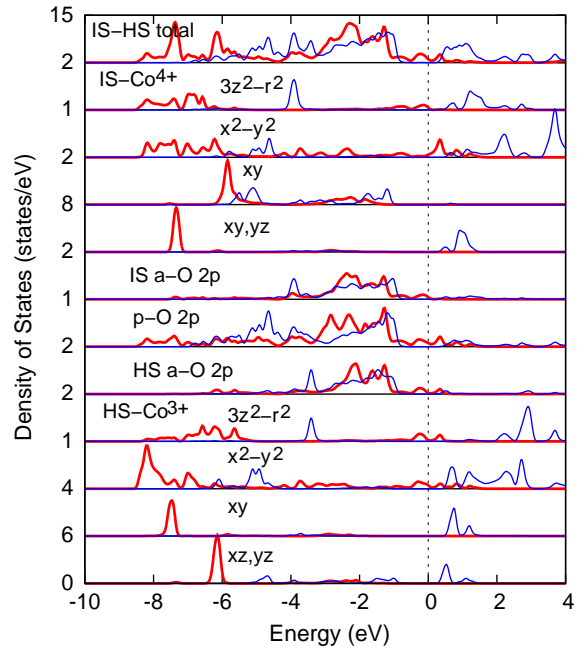} 
\caption{(Color online)
DOS of {\SLCO} in the IS Co$^{4+}$-HS Co$^{3+}$ state calculated by GGA+$U$.
It is FM half metallic.
}
\end{figure}

Taking into account the above crystal-field scenario, we then calculate the LS Co$^{4+}$-HS 
Co$^{3+}$ state, i.e., our calculations now involve a possible spin-state transition of the 
Co$^{4+}$ ion from the IS state in {\SCO} to the LS state in {\SLCO}. [This spin-state 
transition is likely upon La doping into {\SCO}, as the introduced HS Co$^{3+}$ ions get bigger 
in size (compared with the IS Co$^{4+}$ ions), and then a chemical pressure forces
the IS Co$^{4+}$ ions transit into the lower-volume LS state.]    
At a glance, an electron hopping from Co$^{3+}$ to Co$^{4+}$ is possible,
as there could be no change of the electronic configurations (from the initial 
$d^6$+$d^5$ state to the final $d^5$+$d^6$ state). However, as seen in Figs. 1(b) and 1(c), 
the $x^2$--$y^2$ electron hopping from HS Co$^{3+}$ to LS Co$^{4+}$ is actually suppressed, particularly due to the spin-state issue. If such a hopping took place, there would be an energy
cost of Hund exchange associated with the change of the spin states from LS-HS to IS-IS.
As a result, the LS Co$^{4+}$-HS Co$^{3+}$ state will be stabilized at it is, and it has no  
real electron hopping (due to a spin blockade\cite{Chang,Maignan}) and is thus insulating. 
However, a virtual hopping of the $x^2$--$y^2$ 
electron forth and back and the local Hund exchange would mediate a superexchange FM in the
$e_g^0$-$e_g^2$ configuration of the LS Co$^{4+}$-HS Co$^{3+}$ state. 

\begin{figure}[t]
\centering \includegraphics[width=8cm]{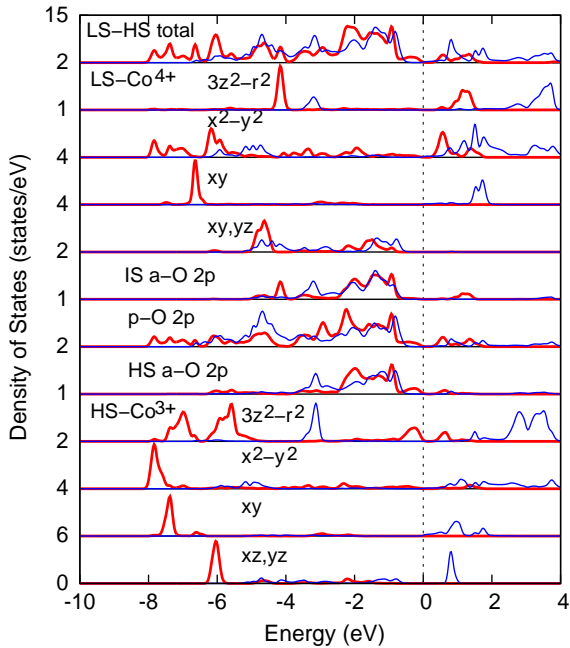} 
\caption{(Color online)
DOS of {\SLCO} in the LS Co$^{4+}$-HS Co$^{3+}$ ground state calculated by GGA+$U$.
It is FM semiconducting.
}
\end{figure}

The insulating but FM solution of the LS Co$^{4+}$-HS Co$^{3+}$ state is indeed confirmed 
by our calculations, 
see Fig. 4: it has a small insulating gap of about 0.3 eV. The Co$^{4+}$ $t_{2g}$ orbitals
carry a spin=1/2 (on the $xy$ orbital), and the strong $pd\sigma$ hybridization brings about 
a large amount of $e_g$ occupation and thus increases the local spin moment of the Co$^{4+}$ 
to 1.7 $\mu_B$, see Table I. The LS Co$^{4+}$ ion has a negligible orbital moment of 0.01 $\mu_B$.
For the Co$^{3+}$ ion, Co-O valency reduces its local spin moment to  
2.88 $\mu_B$. The total spin moment is 5 $\mu_B$/2fu as expected for this LS Co$^{4+}$-HS
Co$^{3+}$ ($S$=1/2 plus $S$=2) state. In this semiconducting state, besides the localized 
$t_{2g}$ and $3z^2$--$r^2$ orbitals,
the $x^2$--$r^2$ orbital also gets localized due to the spin blockade. Then the formal HS
Co$^{3+}$ ion has the configuration state 
$t_{2g\uparrow}^3$$e_{g\uparrow}^2$($xz$+$iyz$)$_{\downarrow}^1$, i.e., the up-spin orbitals
are fully occupied and form a closed subshell, but the rest single electron occupies the 
lowest crystal-field doublet $xz$/$yz$. As a result, the spin-orbit coupling lifts the orbital 
degeneracy, produces the $Y_{21}$ ($xz$+$iyz$) complex orbital, and gives a large orbital moment 
of 0.84 $\mu_B$ along the easy-magnetization $c$-axis. 
The importance of spin-orbit coupling was also demonstrated previously when studing 
the spin-orbital ground state of the perovskite LaCoO$_3$.\cite{Ropka}
Thus, in {\SLCO} the total spin and orbital moments sum up to about 
5.8 $\mu_B$/2fu. 

It is important to note that this LS Co$^{4+}$-HS Co$^{3+}$ 
FM state is more stable than the above IS-HS FM state by 139 meV/fu. 
In addition, this LS-HS state brings about a local distortion against the homogeneous
Co-O lattice measured so far in polycrystals. Our calculations doing atomic relaxations show 
that the LS Co$^{4+}$ ion has the optimized Co-O bondlengths of 1.879 \AA~$\times$ 4 (in plane) 
and 2.011 \AA~$\times$ 2 (out of plane). The corresponding values are 1.911 \AA~$\times$ 4 and 
2.061 \AA~$\times$ 2 for the HS Co$^{3+}$. Such atomic relaxations help the LS Co$^{4+}$-HS 
Co$^{3+}$ state gain further an elastic energy of 54 meV/fu as shown by our calculations.
The semiconducting electronic structure turns out to have insignificant changes due to the atomic 
relaxations. For example, the local spin and orbital moments of the LS Co$^{4+}$ 
(the HS Co$^{3+}$) are now 1.64 and 0.01 (2.92 and 0.83) $\mu_B$, respectively 
(see Table I for a comparison).
Moreover, our calculation finds a LS-HS ferrimagnetic solution to be less stable than the
LS-HS FM solution by 34 meV/fu. The LS-HS ferrimagnetic solution has a total spin moment
of 3 $\mu_B$/2fu as expected. The constituent LS Co$^{4+}$ ($S$ = --1/2) has a local 
spin (orbital) moment of --0.42 (--0.01) $\mu_B$, and the HS Co$^{3+}$ ($S$ = 2) has 
2.97 (0.84) $\mu_B$.  
Furthermore, our calculations show that another possible spin state of {\SLCO}---the IS 
Co$^{4+}$-IS Co$^{3+}$ state is unstable and converges to the present LS-HS
FM state. Therefore, the present FM semiconducting LS Co$^{4+}$-HS Co$^{3+}$ solution is 
the ground state of {\SLCO}. 

This FM semiconducting ground-state solution agrees with the experiments.\cite{Shimada,Chichev}
Moreover, using the LS Co$^{4+}$-HS Co$^{3+}$ state ($S$=$\frac{1}{2}$ and $S$=2 plus $L$=1), 
we estimate the effective magnetic moment
$\mu_{eff}$=$\sqrt{0.5\times4\times\frac{1}{2}\times\frac{3}{2} + 
0.5\times(4\times2\times3+1\times1\times2)}$ $\approx$ 3.8 $\mu_B$/Co. Taking a covalency reduction, this value is in good agreement with the 
measured one of 3.5 $\mu_B$.\cite{Shimada} Although the predicted magnetic moment of 
2.9 $\mu_B$/fu (i.e., the above 5.8 $\mu_B$/2fu) along the easy-magnetization $c$-axis 
(of a single crystal, ideally)
is much bigger than the measured 1.5 $\mu_B$,\cite{Shimada} using of the polycrystals so far 
in the experiments would most probably account for the large reduction. 
The predicted LS-HS state, a big orbital 
moment along the $c$-axis, and the local Co-O distortions in {\SLCO} call for further studies 
on a single crystal. 
  
\section{Conclusion}

In summary, using GGA+$U$ calculations, we find that while {\SCO} in the Co$^{4+}$
IS ground state is a FM half-metal, {\SLCO} has the LS Co$^{4+}$-HS Co$^{3+}$ ground state.
It is the Co$^{4+}$ IS-LS transition
that suppresses an electron hopping via a spin blockade and thus drives 
a metal-insulator transition in {\SLCO}. Moreover,
the present spin-state picture consistently accounts for the FM behavior of
metallic {\SCO} via a $pd\sigma$ hybridization and that of semiconducting {\SLCO} 
via a superexchange. Thus, spin state could provide a way to tune materials properties. 
\\

This work was supported by the NSF of China and WHMFC (WHMFCKF2011008).


\begin{thebibliography}{15}

\bibitem{Tokura} 
M. Imada, A. Fujimori, and Y. Tokura, Rev. Mod. Phys. {\bf 70}, 1039 (1998).

\bibitem{Braden}
A. C. Komarek, S. V. Streltsov, M. Isobe, T. M\"oller, M. Hoelzel, A. Senyshyn, D. Trots, 
M. T. Fern\'andez-D\'iaz, T. Hansen, H. Gotou, T. Yagi, Y. Ueda, V. I. Anisimov, 
M. Gr\"uninger, D. I. Khomskii, and M. Braden,
Phys. Rev. Lett. {\bf 101}, 167204 (2008).

\bibitem{Matsuno}
J. Matsuno, Y. Okimoto, Z. Fang, X. Z. Yu, Y. Matsui, N. Nagaosa, M. Kawasaki, 
and Y. Tokura, Phys. Rev. Lett. {\bf 93}, 167202 (2004).

\bibitem{WangXL}
X. L. Wang and E. Takayama-Muromachi, Phys. Rev. B {\bf 72}, 064401 (2005).

\bibitem{Pandey}
S. K. Pandey, Phys. Rev. B {\bf 81}, 035114 (2010).

\bibitem{Shimada}
Y. Shimada, S. Miyasaka, R. Kumai, and Y. Tokura, Phys. Rev. B {\bf 73}, 134424 (2006).

\bibitem{Chichev}
A. V. Chichev, M. Dlouh\'{a}, S. Vratislav, K. Kn\'{i}\v{z}ek, J. Hejtm\'{a}nek, 
M. Mary\v{s}ko, M. Veverka, Z. Jir\'{a}k, N. O. Golosova, D. P. Kozlenko, and B. N. Savenko, 
Phys. Rev. B {\bf 74}, 134414 (2006).

\bibitem{Wu10}
H. Wu, Phys. Rev. B {\bf 81}, 115127 (2010).

\bibitem{Hollmann08}
N. Hollmann, M. W. Haverkort, M. Cwik, M. Benomar, M. Reuther, A. Tanaka, and T. Lorenz, 
New J. Phys. {\bf 10}, 023018 (2008).

\bibitem{Cwik}
M. Cwik, M. Benomar, T. Finger, Y. Sidis, D. Senff, M. Reuther, T. Lorenz, and M. Braden, 
Phys. Rev. Lett. {\bf 102}, 057201 (2009).

\bibitem{Chang}
C. F. Chang, Z. Hu, H. Wu, T. Burnus, N. Hollmann, M. Benomar, T. Lorenz, A. Tanaka, 
H.-J. Lin, H. H. Hsieh, C. T. Chen, and L. H. Tjeng, 
Phys. Rev. Lett. {\bf 102}, 116401 (2009).

\bibitem{Wu09} H. Wu and T. Burnus, Phys. Rev. B {\bf 80}, 081105(R) (2009). 

\bibitem{Helme}
L. M. Helme, A. T. Boothroyd, R. Coldea, D. Prabhakaran, C. D. Frost, D. A. Keen, 
L. P. Regnault, P. G. Freeman, M. Enderle, and J. Kulda, 
Phys. Rev. B {\bf 80}, 134414 (2009).

\bibitem{Tealdi}
C. Tealdi, C. Ferrara, L. Malavasi, P. Mustarelli, C. Ritter, G. Chiodelli, and 
Y. A. Diaz-Fernandez, Phys. Rev. B {\bf 82}, 174118 (2010). 

\bibitem{Babkevich}
P. Babkevich, D. Prabhakaran, C. D. Frost, and A. T. Boothroyd,
Phys. Rev. B {\bf 82}, 184425 (2010). 

\bibitem{Hollmann11}
N. Hollmann, M. W. Haverkort, M. Benomar, M. Cwik, M. Braden, and T. Lorenz,
Phys. Rev. B {\bf 83}, 174435 (2011). 

\bibitem{Merz}
M. Merz, D. Fuchs, A. Assmann, S. Uebe, H. v. L\"ohneysen, P. Nagel, and S. Schuppler,
Phys. Rev. B {\bf 84}, 014436 (2011). 

\bibitem{Anisimov} V. I. Anisimov, I. V. Solovyev, M. A. Korotin, 
M. T. Czy\.zyk, and G. A. Sawatzky, Phys. Rev. B {\bf 48}, 16929 (1993). 

\bibitem{WangJ}
J. Wang, W. Zhang, and D. Y. Xing, Phys. Rev. B {\bf 62}, 14140 (2000); 
J. Wang, Y. C. Tao, W. Zhang, and D. Y. Xing, 
J. Phys.: Condens. Matter {\bf 12}, 7425 (2000).

\bibitem{Blaha} P. Blaha, K. Schwarz, G. Madsen, D. Kvasnicka, and J. Luitz,
{\bf WIEN2k}, 2001. ISBN 3-9501031-1-2.

\bibitem{Potze} R. H. Potze, G. A. Sawatzky, and M. Abbate,
Phys. Rev. B {\bf 51}, 11501 (1995).

\bibitem{Maignan} A. Maignan, V. Caignaert, B. Raveau, D. Khomskii,
and G. Sawatzky, Phys. Rev. Lett. {\bf 93}, 026401 (2004).

\bibitem{Ropka}
Z. Ropka and R. J. Radwanski, Phys. Rev. B {\bf 67}, 172401 (2003).

\end{thebibliography}
\end{document}